\documentclass[10pt,twocolumn,superscriptaddress,english,10pt,prl,showpacs,floatfix,aps]{revtex4-1}
\usepackage[utf8]{inputenc}
\usepackage{amsthm}
\usepackage{amsmath}
\usepackage{amssymb}
\usepackage{esint}
\usepackage{graphicx}
\usepackage{upgreek}
\usepackage{braket}

\usepackage[markup=nocolor%
						,deletedmarkup=xout%
						]{changes}
\usepackage{xspace}

\makeatletter
\@ifundefined{textcolor}{}
{%
 \definecolor{BLACK}{gray}{0}
 \definecolor{WHITE}{gray}{1}
 \definecolor{RED}{rgb}{1,0,0}
 \definecolor{GREEN}{rgb}{0,1,0}
 \definecolor{BLUE}{rgb}{0,0,1}
 \definecolor{CYAN}{cmyk}{1,0,0,0}
 \definecolor{MAGENTA}{cmyk}{0,1,0,0}
 \definecolor{YELLOW}{cmyk}{0,0,1,0}
}

\usepackage{soul}
\usepackage[backref=none,
bookmarksnumbered=true,
bookmarks=true,
bookmarksopen=true,
colorlinks=true,
citecolor=blue,
linkcolor=blue,
anchorcolor=green,
urlcolor=blue,unicode=false]{hyperref}

\let\baraccent=\= 
\renewcommand{\=}[1]{\stackrel{#1}{=}} 

\newcommand{\unitspace}{~}

\newcommand{\didv}{\ensuremath{{\rm d}I/{\rm d}V}\xspace}

\newcommand{\DeltaT}{\ensuremath{\Delta_\mathrm{tip}}\xspace}
\newcommand{\DeltaS}{\ensuremath{\Delta_\mathrm{sample}}\xspace}

\newcommand{\Fig}[1]{\mbox{Fig.\unitspace\ref{fig:#1}}}
\newcommand{\Figs}[1]{\mbox{Figs.\unitspace\ref{fig:#1}}}
\newcommand{\Figure}[1]{\mbox{Figure\unitspace\ref{fig:#1}}}

\newcommand{\Refs}[1]{Refs.~\cite{#1}}

\newcommand{\YSR}{YSR\xspace}

\newcommand{\dir}[1]{$\left[#1\right]$}
\newcommand{\dorb}[1]{$d_\mathrm{#1}$}

\DeclareMathOperator{\unm}{\unitspace\mathrm{nm}}
\DeclareMathOperator{\upm}{\unitspace\mathrm{pm}}
\DeclareMathOperator{\uV}{\unitspace\mathrm{V}}
\DeclareMathOperator{\umV}{\unitspace\mathrm{mV}}
\DeclareMathOperator{\umuV}{\unitspace\mathrm{\mu V}}
\DeclareMathOperator{\ueV}{\unitspace\mathrm{eV}}
\DeclareMathOperator{\umeV}{\unitspace\mathrm{meV}}
\DeclareMathOperator{\umueV}{\unitspace\mathrm{\mu eV}}

\DeclareMathOperator{\upA}{\unitspace\mathrm{pA}}
\DeclareMathOperator{\unS}{\unitspace\mathrm{nS}}
\DeclareMathOperator{\umuS}{\unitspace\mathrm{\mu S}}
\DeclareMathOperator{\uK}{\unitspace\mathrm{K}}

\DeclareMathOperator{\uHz}{\unitspace\mathrm{Hz}}

\DeclareMathOperator{\uT}{\unitspace\mathrm{T}}
\DeclareMathOperator{\umT}{\unitspace\mathrm{mT}}
\DeclareMathOperator{\uAA}{\unitspace\mathrm{\AA{}}}

\makeatother

\usepackage{babel}

\begin{document}

\author{Michael Ruby}
\affiliation{Fachbereich Physik, Freie Universit\"at Berlin, 14195 Berlin, Germany}
\author{Benjamin W. Heinrich}\email{bheinrich@physik.fu-berlin.de}

\affiliation{Fachbereich Physik, Freie Universit\"at Berlin, 14195 Berlin, Germany}

\author{Yang Peng}
\affiliation{Fachbereich Physik, Freie Universit\"at Berlin, 14195 Berlin, Germany}
\affiliation{{Dahlem Center for Complex Quantum Systems, Freie Universit\"at Berlin, 14195 Berlin, Germany}}

\author{Felix von Oppen}
\affiliation{Fachbereich Physik, Freie Universit\"at Berlin, 14195 Berlin, Germany} 
\affiliation{{Dahlem Center for Complex Quantum Systems, Freie Universit\"at Berlin, 14195 Berlin, Germany}}

\author{Katharina J. Franke}
\affiliation{Fachbereich Physik, Freie Universit\"at Berlin, 14195 Berlin, Germany}
\date{\today}

\title{Exploring a proximity-coupled Co chain on Pb(110) as a possible Majorana platform}
\keywords{Yu-Shiba-Rusinov state, spin-polarization, topological superconductivity, Majorana fermion, proximity-coupled chain}

\begin{abstract}
Linear, suspended chains of magnetic atoms proximity coupled to an $s$-wave superconductor are predicted to host Majorana zero modes at the chain ends in the presence of strong spin-orbit coupling. Specifically, iron (Fe) chains on Pb(110) have been explored as a possible system to exhibit topological superconductivity and host Majorana zero-modes [Nadj-Perge, et al., Science 346, 602 (2014)]. Here, we study chains of the transition metal cobalt (Co) on Pb(110) and check for topological signatures.
Using spin-polarized scanning tunneling spectroscopy, we resolve ferromagnetic order in the $d$ bands of the chains. Interestingly, also the subgap Yu-Shiba-Rusinov (\YSR) bands carry a spin polarization as was predicted decades ago. Superconducting tips allow us to resolve further details of the \YSR bands and in particular resonances at zero energy. We map the spatial distribution of the zero-energy signal and find it delocalized along the chain. Hence, despite of the ferromagnetic coupling within the chains and the strong-spin orbit coupling in the superconductor, we do not find clear evidence of Majorana modes. Simple tight-binding calculations suggest that the spin-orbit-split bands may cross the Fermi level four times which suppresses the zero-energy modes.
\end{abstract}

\maketitle
Low-dimensional structures proximity coupled to an s-wave superconductor can support topologically protected Majorana zero modes, which obey non-Abelian statistics and are potentially useful for fault-tolerant quantum computation~\cite{Alicea12,Beenakker13,Elliott15}. Although realizing a topological superconductor is challenging, there are promising results for various experimental platforms~\cite{mourik12,Nadj14, albrecht16,lv16,menard16}. 
The simplest systems emulate a model first proposed by Kitaev~\cite{kitaev01}: a tight-binding chain for a spinless $p$-wave superconductor in one dimension. This system with nearest-neighbor hopping and pairing carries zero energy excitations at the chain ends which are protected when an odd number of bands cross the Fermi level. Ferromagnetic chains of atoms adsorbed on an $s$-wave superconductor in the presence of strong spin-orbit coupling have been suggested as an intriguingly simple experimental realization~\cite{Li14}. By proximity coupling Cooper pairs enter the chain, which induces $p$-wave superconductivity in the chain and turns it into a topological superconductor. 
The occurrence of Majorana modes has been predicted to be near universal in sufficiently long chains of transition-metal atoms on Pb \cite{Li14}.
Such systems fulfill the requirements for a topological superconducting phase: a large exchange splitting of the $d$ bands, strong Rashba spin-orbit coupling originating from the superconducting substrate, and proximity induced superconductivity. 

Nadj-Perge et al. have presented indications of Majorana zero modes in Iron (Fe) chains on Pb(110) ~\cite{Nadj14}. Additional experiments have stimulated further discussions ~\cite{pawlak15,RubyMaj15,Feldman16}. 
The appealing simplicity of this platform motivates us to explore chains of another $3d$ element. We replace iron by cobalt (Co), thereby keeping the one-dimensional band structure of the chain similar while modifying the number of $d$ electrons and hence the band filling.




\begin{figure}[bt]
	\includegraphics[width=\columnwidth]{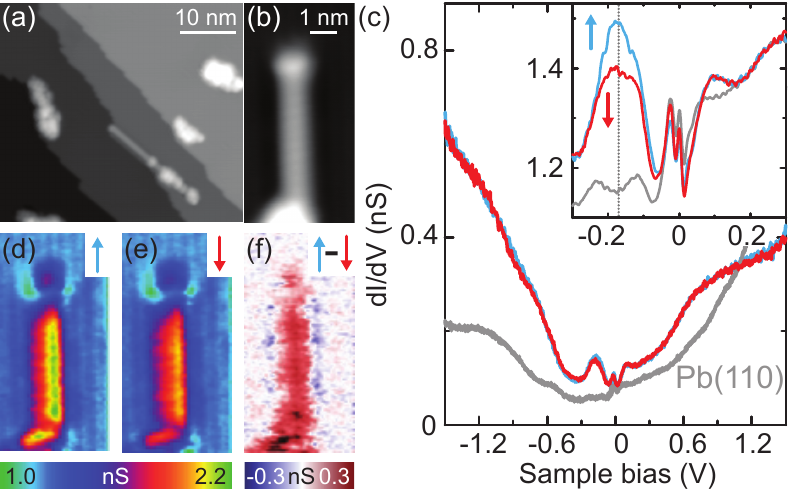}
	\caption{Co chain on Pb(110). (a) After deposition on Pb(110), Co forms clusters and 1D chains. Setpoint: $100\umV, 50\upA$. (b) Closeup of a 1D Co chain of length $\approx5.5\unm$. Setpoint: $500\umV, 50\upA$. (c) Spin-polarized \didv spectra of the chain shown in (b) ({\it blue, red}) and of pristine Pb(110) (\textit{gray}). The \textit{blue} (\textit{red}) curve is acquired in a $+(-)0.3\uT$ magnetic field along the surface normal. The field reversal ensures opposite tip magnetization (in the following $\uparrow$ and $\downarrow$). Setpoint: $1.5\uV$, $400\upA$. The inset shows spectra in a narrower energy window. Setpoint: $300\umV$, $400\upA$. (d) and (e) show \didv maps at $V=-170\umV$ in fields of $+$ and $-0.3\uT$, respectively, revealing spin-dependent \didv intensities all along the chain. Feedback: $300\umV$, $400\upA$. (f) Map of the signal difference of (d) and (e).}
\label{fig:SpinPolDBands}
\end{figure}

At a sample temperature of $263\uK$, Co deposition onto Pb(110) yields clusters and 1D chains with lengths of up to $\simeq11\unm$ [\Fig{SpinPolDBands}(a)]. These resemble the Fe chains studied earlier~\cite{Nadj14,pawlak15,RubyMaj15,Feldman16}. In most cases, the chains emerge from a Co cluster and follow the \dir{1\bar{1}0} direction of the (110) surface. At the opposite end, the chains are either flat or terminated by a small protrusion as was also observed in the case of Fe. \Figure{SpinPolDBands}(b) presents a closeup of a typical chain of $\approx5.5\unm$ length (measured between chain end and the onset of the cluster).

To probe the magnetic properties of the chain~\cite{Wiesend2009}, we employ Co-coated tips, which have been tested for their magnetization beforehand (see Methods and Supporting Information~\cite{supplementary}). We resolve a resonance at $-0.17\uV$ [\Fig{SpinPolDBands}(c)], which exhibits different intensities for oppositely polarized tips (labeled $\uparrow$ and $\downarrow$). We ascribe this resonance to the van Hove singularity of a spin-polarized Co $d$ band (for additional spectra along the same chain, see the Supporting Information~\cite{supplementary}). The magnetic order of the chain is revealed by \didv maps at the energy of the van Hove singularity [\Fig{SpinPolDBands}(d,e)]. The intensity along the chain is stronger for tip$_\uparrow$ than for tip$_\downarrow$. In the difference map shown in \Fig{SpinPolDBands}(f), this leads to a positive contrast ({\it red}) on the chain. The uniform contrast along the Co chain suggests that it is in a ferromagnetic state, similar to Fe chains on Pb(110). 

\begin{figure}[tb]
	\includegraphics[width=\columnwidth]{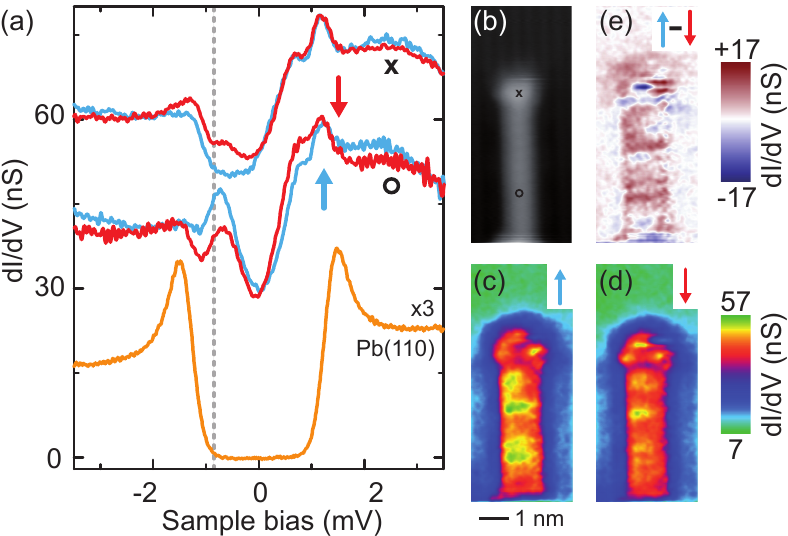}
	\caption{(a) \didv spectra taken at the two positions on the Co chain depicted in the topography in (b).  For clarity, the spectrum taken on the bare surface is divided by three. The data is recorded at $0\uT$ with a spin-polarized tip with a remanence which has an out-of-plane component relative to the sample surface. Parallel and anti-parallel orientations are indicated by $\uparrow$ (\textit{blue}) and $\downarrow$ (\textit{red}). Setpoint: $4\umV$, $200\upA$. The dashed line at $-850\umuV$ marks, where the \didv maps in (c,d) are recorded with parallel and anti-parallel tip-magnetization, respectively. The feedback was opened at $4\umeV$, $200\upA$ where no spin-polarization is observed. (e) Difference between (c) and (d) characterizing the spin-polarization of the chain.}
	\label{fig:SpinPolShibas}
\end{figure}

Figure \ref{fig:SpinPolShibas} explores magnetic signatures within the superconducting energy gap using the same tip as before.  The \didv spectrum on pristine Pb(110) shows a BCS-like gap, broadened by the Fermi-Dirac distribution of the tip at 1.1\,K [\Fig{SpinPolShibas}(a)]. On the chain, there are broad resonances within the gap, which vary in intensity along the chain (see the Supporting Information~\cite{supplementary} for additional spectra). These resonances reflect Yu-Shiba-Rusinov~\cite{yu,shiba,rusinov} (\YSR) bound states, which result from the exchange coupling between the spin-polarized Co $d$ states and the superconducting substrate~\cite{yu,shiba,rusinov}. When measured with opposite tip magnetization (which is possible in zero field because of a sizable magnetic remanence of the tip~\cite{supplementary}), the spectra are qualitatively similar, but differ in signal strength. The overall intensity at negative (positive) energies is stronger (weaker) for tip$_\uparrow$ than for  tip$_\downarrow$. This is clearly revealed by the spin contrast map [\Fig{SpinPolShibas}(e)], which exhibits an overall positive polarization along the chain at $-850\,\umuV$, but a negative polarization at $+850\,\umuV$ (see the Supporting Information~\cite{supplementary}). 
This sizable spin polarization is remarkable because it provides direct experimental evidence for the magnetic nature of \YSR bands, which was predicted theoretically decades ago~\cite{yu,shiba,rusinov}. The hybridization of \YSR states of neighboring adatoms along the chain results in spin-polarized bands. Although confinement effects and potential variations cause intensity variations of the \YSR bands \cite{RubyMaj15}, the spin-polarization is almost uniform (at $-850\,\umuV$). Only at the chain end, a region of opposite polarization is detected. However, the magnetization of the chains presumably does not change sign as indicated by the uniform polarization of the $d$ bands [see \Fig{SpinPolDBands}(f)].

\begin{figure}[tb]
	\includegraphics[width=\columnwidth]{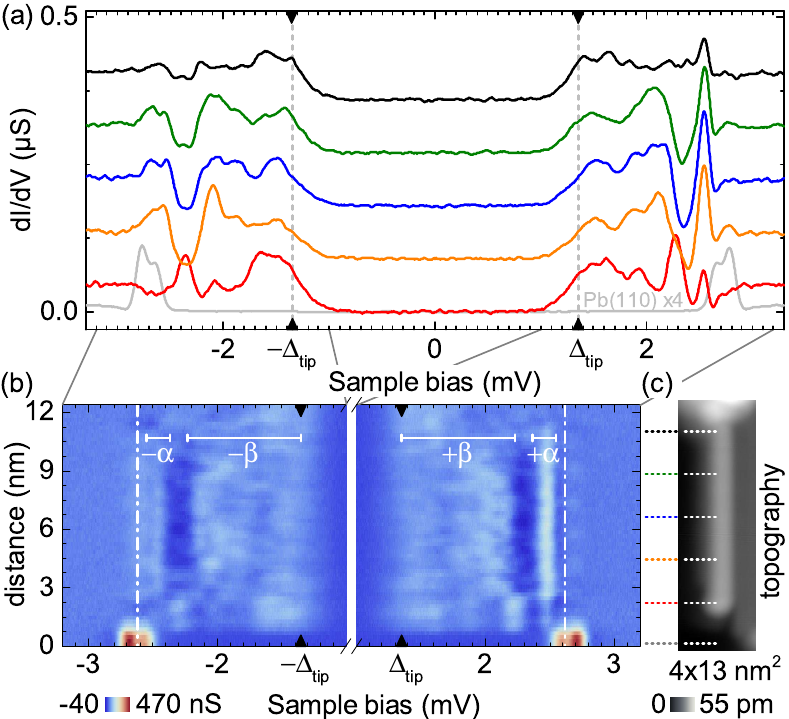}
	\caption{(a) \didv spectra acquired with a superconducting tip on the bare surface (\textit{gray}; divided by four) and on the $\simeq10.3\unm$ long chain shown in the topography in (c). Spectra offset by $90\unS$ for clarity, tip positions indicated by colored dashed lines in (c). The tip gap is marked by dashed lines ($\pm\DeltaT=\pm1.35\umeV$). Setpoint: $5\umV$, $200\upA$. (b) False-color plot of all 40 \didv spectra measured along the central axis of the chain in (c). Beside an intense resonance close to the gap edge [$\alpha\simeq(2.5\pm0.1)\umV$], spectral intensity appears mainly in the energy interval $\beta\simeq(1.8\pm0.4)\umV$.
As a guide to the eye, the dashed-dotted lines indicate the gap edge at $eV=\pm(\DeltaT+\DeltaS)$. See Fig.~S6 and S7 of the Supporting Information for additional data on this chain~\cite{supplementary}. }
	\label{fig:SCShibasLineScan}
	\label{fig:SCShibas}
\end{figure}

\begin{figure}[tb]
	\includegraphics[width=\columnwidth]{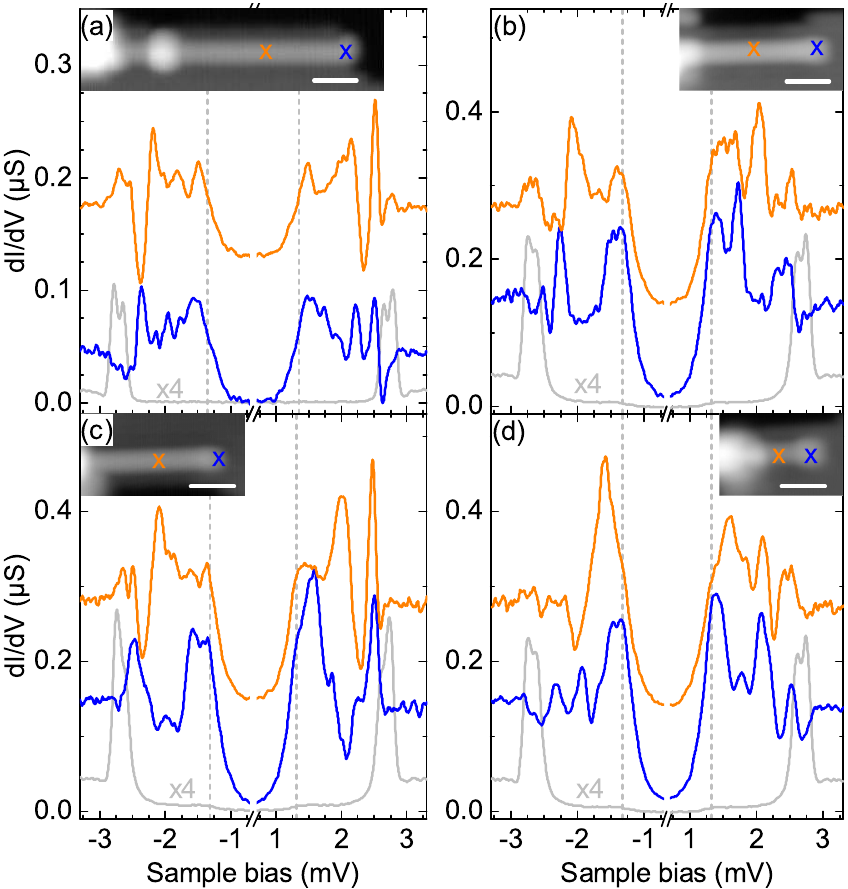}
	\caption{\didv spectra acquired at the end and center, respectively, of four chains with length ranging from $2.5$ to $11.7\unm$ (tip positions marked in the corresponding topographies). The spectra were recorded with different tips with superconducting gaps of $1.35\umeV$ for (a), and of $1.32\umeV$ for (b--d), respectively. As guide to the eye the energy of the tip gap is marked by dashed lines. A spectrum of the bare surface is superimposed in gray for comparison and divided by four. The scale bars in the inset correspond to $2\unm$. Spectra are offset by $0.13\umuS$ for clarity. See Fig.~S8 to S10 of the Supporting Information for more spectra of these chains~\cite{supplementary}.} 
	\label{fig:SCShibasMultiple}
\end{figure}

To gain more detailed insight into the quasiparticle excitations and to explore the possibility of Majorana zero modes at the chain ends, we use a superconducting Pb tip (\Fig{SCShibas}). This increases the energy resolution well beyond the Fermi-Dirac limit, but shifts all spectral features by $\DeltaT$, the gap of the superconducting tip. Putative Majorana modes should thus appear at $eV=\pm\DeltaT$. On Pb(110) [\Fig{SCShibas}(a), {\it grey}], we resolve the double peak structure of the coherence peaks of the two-band superconductor Pb at $\pm(\DeltaT+\DeltaS)$~\cite{ruby15a}. On the chains, we find a rich subgap structure, which varies along the chains [see \Figs{SCShibas}(a) and (b) for a chain of  $\simeq10.3\unm$]. The most intense resonance resides close to the superconducting gap edge [labeled $\alpha$ \Fig{SCShibasLineScan}(b)] and is well separated from a broader band of resonances at lower energy [$\beta$ in \Fig{SCShibasLineScan}(b)]. The pronounced separation of $\alpha$ and $\beta$ is observed for most, but not all of the chains investigated (see \Fig{SCShibasMultiple}, as well as Fig.~S4 and S5 in the Supporting Information for data on additional chains~\cite{supplementary}).

The resonances persist throughout the chain, but show local intensity variations arising from confinement effects and variations in the local potential~\cite{RubyMaj15}. At zero energy, i.e., at a bias voltage $\simeq\pm\DeltaT$, we observe resonances (or shoulders), which might at first sight be reminiscent of Majorana states. However, the zero-energy signal is present all along the chain with no sign of localization at the chain end, in contrast to the expected signature for Majorana zero modes. 
   
Next, we explore the influence of the chain length on the excitation spectrum in \Fig{SCShibasMultiple} (data on additional chains with different lengths are shown in the Supporting Information~\cite{supplementary}).  For all chains, \didv spectra acquired at the end exhibit a rich subgap structure and a sizable spectral intensity at $\simeq\pm\DeltaT$.  However, similar spectral intensity is also present in  spectra recorded in the center of the chains (in agreement with the chain presented above). We investigated $23$ chains with lengths ranging from $2.5$ to $11.7\unm$. None of them showed localization of zero-energy resonances at the chain end. One might argue that the distance between the end states is within the Majorana localization length. The states would then hybridize and lose their Majorana character. The splitting should be more pronounced the stronger the overlap, that is, the shorter the chain. However, we do not observe any distance dependence. Furthermore, the localization length of Majorana states is expected to be on the order of atomic distances~\cite{peng2015}. The absence of localization in any of the chains suggests that the zero-energy features cannot be assigned to a Majorana mode. 

\begin{figure}[bt]
	\includegraphics[width=\columnwidth]{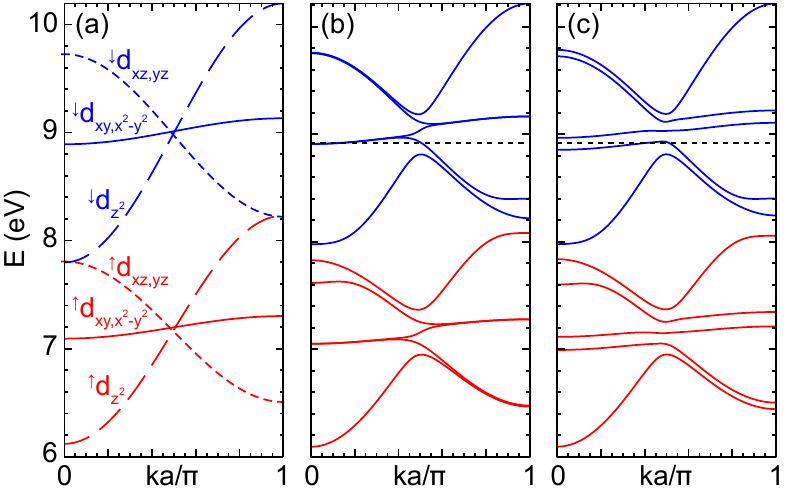}
	\caption{Tight-binding band structures of a linear Co chain with interatomic distance $a=2.486\uAA$ and without coupling to the Pb substrate. Panels (a) and (b,c) show results without and with spin-orbit coupling, respectively. The spin-orbit coupling parameter is $\lambda_\mathrm{so}=0.2\ueV$. In panel (b), the adatom magnetization is perpendicular to the chain direction. In panel (c), the angle between magnetization and chain direction is taken as $2\pi/5$. The exchange interaction splits minority ({\it blue}) and majority ({\it red}) bands. The black dotted lines indicate the chemical potential, which implies three Fermi points within half the Brillouin zone in (b), but two in (c). }
	\label{fig:BandStructureSOC}
\end{figure}

These experimental results indicate the absence of Majorana states although crucial ingredients for topological superconductivity and Majorana zero modes are fulfilled in our system: a ferromagnetic chain with spin-polarized bands at the Fermi level is coupled to an $s$-wave superconductor with strong spin-orbit coupling. 

For a theoretical interpretation, we model the band structure of the linear Co chain. Following calculations for Fe chains \cite{Nadj14}, we employ bulk tight-binding parameters within a Slater-Koster tight-binding approximation for the chain. Moreover, we estimate the filling of the $d$ bands based on the number of $d$ electrons in individual atoms. The $s$~bands are higher in energy and thus have no significant overlap with the $d$~bands. Because of the tendency to form a 2+ or 3+ oxidation state, hybridization with the Pb bulk bands might empty the Co $s$ orbitals (for a more extensive discussion, see the Supporting Information~\cite{supplementary}).
First neglecting spin-orbit coupling [\Fig{BandStructureSOC}(a)], two broad minority bands cross the Fermi energy $E_\mathrm{F}$, one of which is twofold degenerate. In addition, a narrow, doubly degenerate band originating from the \dorb{xy,x^2-y^2} orbitals lies close to the Fermi level. This band might give rise to the resonance at $-0.17\umV$ which we observe in the \didv spectra on the Co chains. When including spin-orbit coupling, all $d$~band degeneracies are lifted. Depending on the relative direction of magnetization and chain, the bands are mixed and shift in energy. We find that unless the magnetization is perpendicular to the chain direction [\Fig{BandStructureSOC}(b)], the system has an even number of Fermi points within half the Brillouin zone.  Thus, even if most prerequisites for the formation of Majorana modes are fulfilled, the hybridization between these Fermi points would prevent the formation of a topological phase. In the case of the Fe chain, the adatoms have one less $d$ electron, resulting in a correspondingly lower $E_\mathrm{F}$. In this case, there are three Fermi points, which would allow Majorana modes. We note, however, that this conclusion is more robust to changes in $E_\mathrm{F}$ for Fe than for Co.

We finally comment on our observation of spectral weight in \didv  at zero energy. This might be a consequence of the subgap band structure. It is possible that the induced gap is below the experimental energy resolution of $\approx60\umuV$, the coherence peaks associated with the gap edges would then not be fully resolved and instead show as spectral weight at zero energy. Clearly, the coherence peaks are a bulk feature of the chain and the corresponding peaks in \didv should persist along the entire chain.

Motivated by the predictions of topological superconductivity as a near universal feature in ferromagnetic chains on superconducting Pb, we deposited Co on Pb(110). 
Similar to Fe~\cite{Nadj14,pawlak15,RubyMaj15}, Co forms one-dimensional chains with ferromagnetic order as evidenced by a homogeneous spin polarization of the $d$ bands. Furthermore, we resolved the spin-polarized nature of \YSR bands with the perspective to probe the polarization of possible Majorana states \cite{Sticlet2012,Bjornson2015} in experiments at lower temperatures. We observed zero-energy spectral weight along the entire chains, albeit without a clear signature of localization at the chain ends, suggesting the absence of topological superconductivity. A simple model of the one-dimensional band structure of the transition metal chains predicts an even number of Fermi points for Co, but a robust topological phase for Fe chains. This highlights the importance of the proper adjustment of the chemical potential to obtain a topologically non-trivial phase. Our work shows that it is rewarding to explore different adatom species as well as superconducting substrates to gain a deeper understanding of topological superconductivity in adatom chains.

\medskip
We gratefully acknowledge funding by the Deutsche Forschungsgemeinschaft through Collaborative Research Centers Sfb 658 and CRC 183, and through Grants FR2726/4 and HE7368/2, as well by the European Research Council through Consolidator Grant NanoSpin. We thank M.\ Font Gual and L.-M. R\"utten for assistance.

\section{Methods}

The experiments are performed in a \textsc{Specs} JT-STM at a temperature of $1.1\uK$ under UHV conditions. The Pb(110) single crystal ($T_\textrm{c}=7.2\uK$) is cleaned by cycles of sputtering and annealing until atomically flat and clean terraces are observed. Co chains were prepared by e-beam evaporation from a cobalt rod (99.995\% purity) onto the clean surface at $263\uK$. Cobalt-covered W-tips are used for spin-polarized measurements. The spin-sensitivity was checked prior to the measurement on bilayer cobalt islands on Cu(111), which posses an out-of-plane magnetization and represent a standard reference system~\cite{Pietzsch04}. The hysteresis loop in an out-of-plane magnetic field reveals a sizeable tip remanence at zero field and a coercivity of $\approx50\umT$~(see the Supporting Information~\cite{supplementary}).
Pb-covered, superconducting tips~\cite{franke11} are used to provide an energy resolution of $\simeq60\umueV$, well beyond the Fermi-Dirac limit. The differential conductance \didv as a function of sample bias was recorded using standard lock-in technique at $912\uHz$ with a bias modulation of $V_\textrm{mod}=15\umuV_\mathrm{rms}$ (Pb tip, $\pm4\umV$), $50\umuV_\mathrm{rms}$ (Co tip, $\pm4\umV$), $5\umV_\mathrm{rms}$ ($\pm0.3\uV$), and $10\umV_\mathrm{rms}$ ($\pm1.5\uV$), respectively.

\clearpage

\setcounter{figure}{0}
\setcounter{section}{0}
\setcounter{equation}{0}
\renewcommand{\theequation}{S\arabic{equation}}
\renewcommand{\thefigure}{S\arabic{figure}}

\onecolumngrid

\renewcommand{\Fig}[1]{\mbox{Fig.\unitspace\ref{Sfig:#1}}}
\renewcommand{\Figs}[1]{\mbox{Figs.\unitspace\ref{Sfig:#1}}}
\renewcommand{\Figure}[1]{\mbox{Figure\unitspace\ref{Sfig:#1}}}

\section*{\Large{Supplementary Material}}

\vspace{0.7cm}

\section{Spin-polarized measurements}
As described in the main manuscript, we preformed spin-polarized measurements~\cite{Wiesend2009} with a cobalt-covered tungsten tip. The tip was cleaned in situ by Ne$^+$ ion sputtering and then coated with two to four layers of cobalt (Co) by e-beam evaporation from a rod. The out-of-plane spin sensitivity of the tip was determined by measurements on the well-established system of two monolayer-high Co islands on Cu(111)~\cite{Pietzsch04}, which posses an out-of-plane magnetization at low temperatures. We characterized the magnetic remanence of the tip by measuring the \didv signal as a function of applied magnetic field. A sizable remanence is essential for measuring spin contrast at zero field. All employed tips showed remanence with a  coercivity of approximately $\pm50\umT$ [\Fig{TipMagnetization}(c)].

\subsection{Spin contrast along the chain and hysteresis of the tip magnetization}

\begin{figure}[b]
	\includegraphics[width=1\columnwidth]{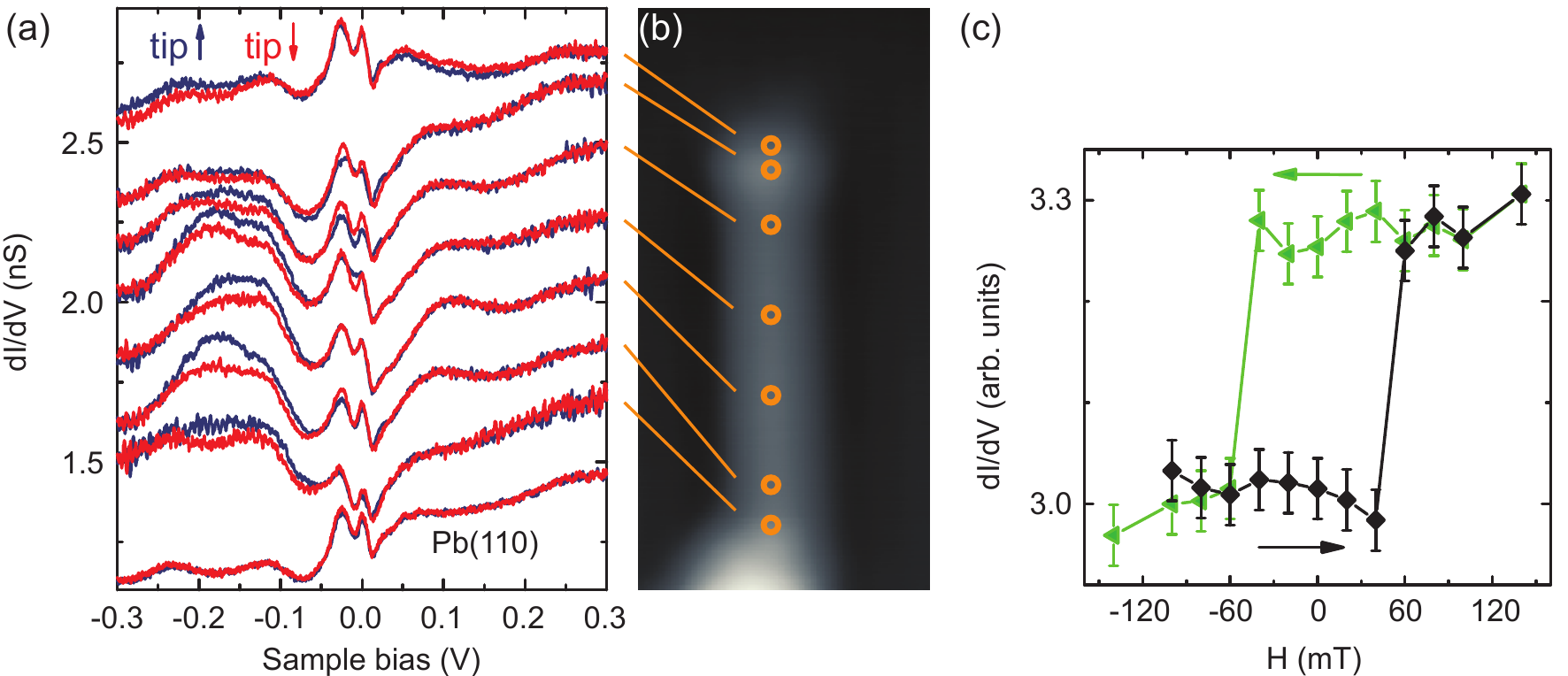}
	\caption{(a) Spin-polarized \didv spectra acquired at different points along the Co chain as indicated in the topography in (b) (same chain as in Fig.~1 and 2 of the main manuscript; $z$ component of the tip magnetization indicated by arrows). Setpoint: $300\umV$, $400\upA$. Bias modulation: $5\umV_\mathrm{rms}$. (c) \didv intensity at $-170\umV$ as a function of the magnetic field strength perpendicular to the sample surface. Field sweep direction is indicated by arrows. Each data point is an  average of the \didv intensity over the same area ($1.5\unm^2$) of the chain.}
	\label{Sfig:TipMagnetization}
\end{figure}

In Fig.~1 of the main text, we plot a \didv spectrum that shows a spin-polarized resonance which we link to a van Hove singularity of a $d$ band of the Co chain on Pb(110). For completeness, \Fig{TipMagnetization}(a) shows a larger set of spectra along the same chain. The spectral intensity of the $d$ band at $-170\umV$ decreases when approaching the chain end or the cluster. However, a spin contrast is detected all along the chain. 

\Figure{TipMagnetization}(c) shows the \didv signal at $-170\umV$ as a function of out-of-plane magnetic field. The data points are obtained as the spatial average of the signal intensity in \didv maps on the chain at the respective field.  A spin contrast is observed  with a hysteresis loop opening around zero field. 
This is due to remanence of the tip magnetization and a coercivity of approximately $\pm50\umT$. The switching of the tip magnetization is in line with the pre-characterization of the tip on Co islands on Cu(111) (not shown).  The magnetization of the chain could not be switched in out-of-plane fields of up to $\pm3\uT$.

\subsection{Spin-polarized excitation spectra of the Yu-Shiba-Rusinov bands}

\begin{figure}[bth]
	\includegraphics[width=1\columnwidth]{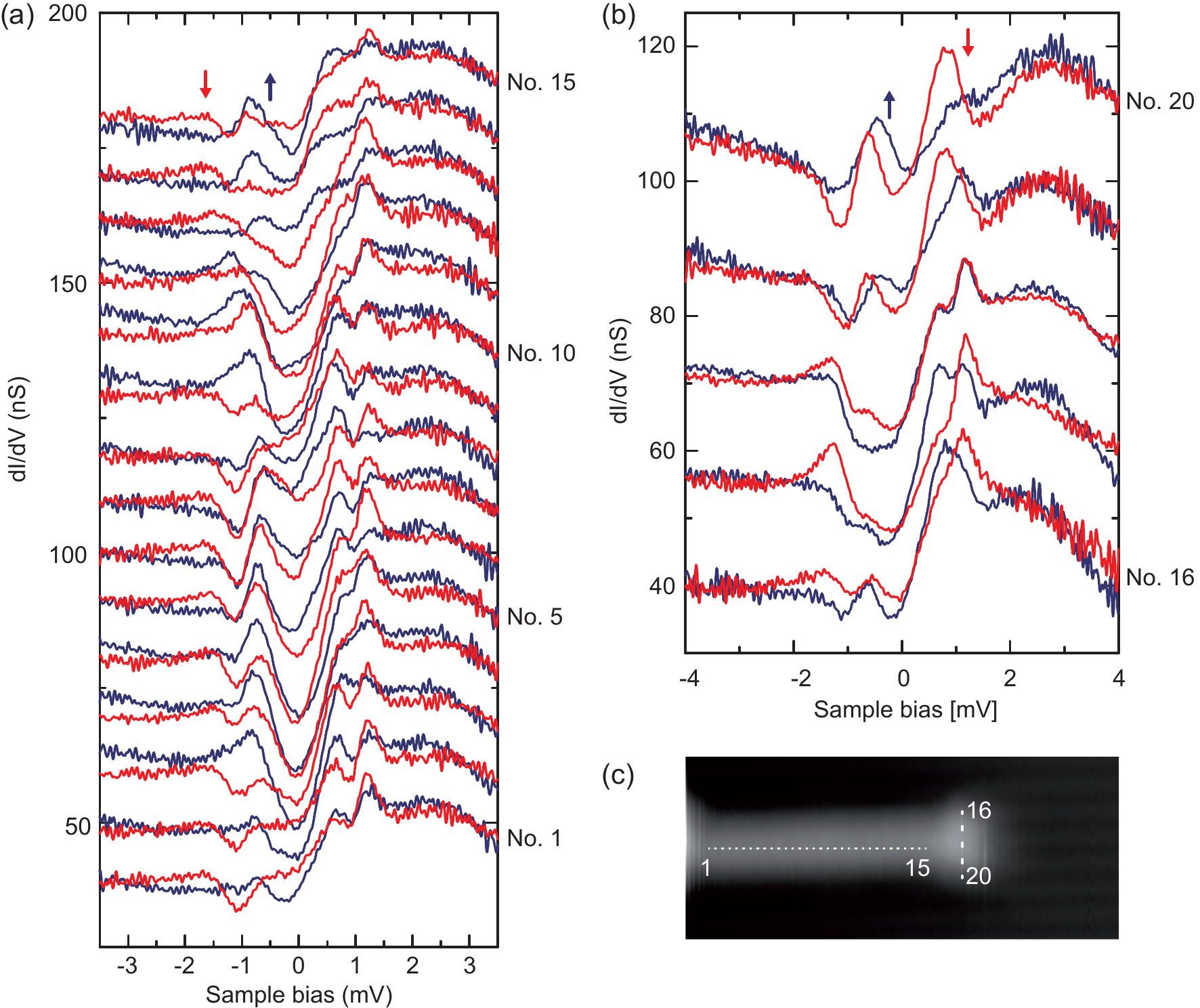}
	\caption{(a,b) Spin-polarized \didv spectra acquired along the body and across the chain end, as indicated in (c). Arrows indicate the orientation of the $z$ component of the tip magnetization. Setpoint: $200\upA, 4\umV$. Bias modulation: $50\umuV_\mathrm{rms}$.}
	\label{Sfig:ShibaLines}
\end{figure}

In Fig.~2 of the main manuscript we showed spin-polarized \didv spectra of the Yu-Shiba-Rusinov (\YSR) bands at the center and at the end of a Co chain. For completeness, we show a set of spectra acquired along the chain's central axis (spectra 1 to 15), and across the chain end (spectra 16 to 20) in \Figs{ShibaLines}(a) and (b), respectively. The signal intensities vary along the chain. However, as shown in the main text and in the next section, the difference maps are rather uniform, besides at the chain end.

\newpage

\subsection{Energy-dependent spin contrast of the Yu-Shiba-Rusinov bands}

\begin{figure}[hbt]
	\includegraphics[width=\columnwidth]{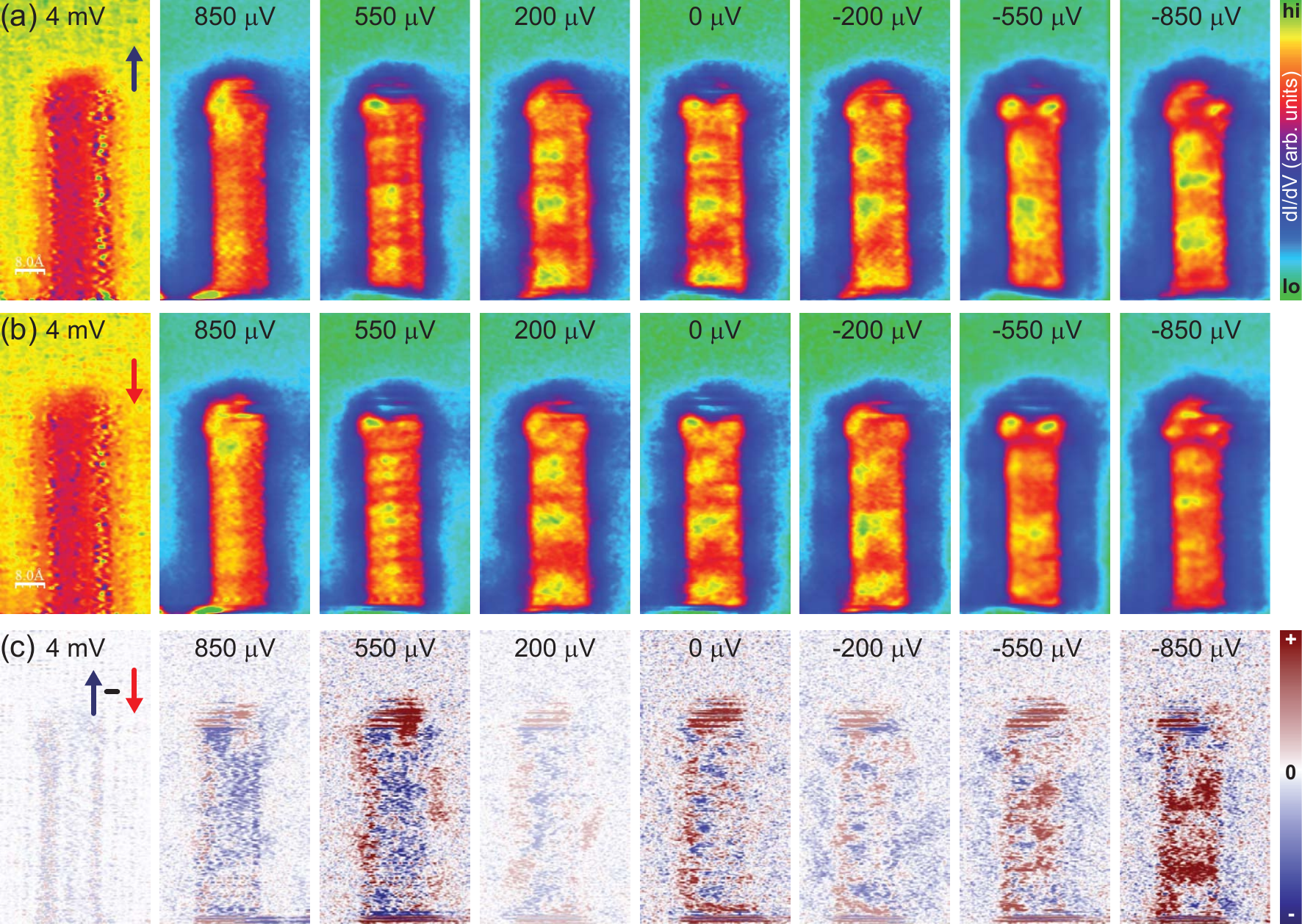}
	\caption{Spin-polarized \didv maps at various subgap energies. Same chain as in Fig.~1 and 2 of the main manuscript. The $z$ component of the tip magnetization is either up (a, tip$_\uparrow$) or down (b, tip$_\downarrow$). (c) Spin contrast maps obtained by subtraction of the respective maps in (a) and (b). Setpoint: $4\umV$, $200\upA$. Bias modulation: $50\umuV_\mathrm{rms}$.}
	\label{Sfig:Shibamaps}
\end{figure}

The main text shows \didv maps of the spectral intensity along the chain at $-850\umuV$ acquired with tip$_\uparrow$ and tip$_\downarrow$, as well as a map of the signal difference in each point (spin contrast map). In \Fig{Shibamaps}, we show \didv maps at other subgap energies, as well as at $4\umV$, i.e., outside the gap, for comparison. All maps reveal local variations of the spectral intensity along the chain. However, the contrast maps at $\pm850\umuV$ and $\pm550\umuV$  unveil a spin contrast along the chain, which is opposite at energies of opposite sign  [\Fig{Shibamaps}(c)].

\newpage

\section{Determination of the superconducting tip gap}

In order to improve the energy resolution beyond the Fermi-Dirac limit we employ superconducting tips for experiments examining the subgap structure.
The \didv spectra result form a convolution of the spectral intensity of the sample with the BCS-like density of states of the tip. One consequence of a superconducting tip with gap \DeltaT is a shift of a sample resonance with energy $\varepsilon$ to a bias value of $\pm\left(\DeltaT +\varepsilon\right)/e$. 
The exact determination of $\varepsilon$ relies on the correct determination of \DeltaT.

Pb is a two-band superconductor with two gap parameters ($\Delta_1\simeq1.42\umeV$ and $\Delta_2\simeq1.29\umeV$). They originate from two separated Fermi surfaces, and give rise to the double-peak structure in the \didv spectra~\cite{ruby15a}. The tip is prepared by controlled indentation into the clean Pb surface with high voltage applied to the tip. This creates an amorphous superconducting Pb layer on the tip and yields a single gap parameter \DeltaT, which can be similar to or smaller than the bulk gap values, depending on the layer thickness and quality.

A spectrum acquired with such a superconducting tip on pristine Pb(110) determines the sums $\Delta_1+$\DeltaT and $\Delta_2+\DeltaT$. Yet, an independent determination of \DeltaT is not straight forward. We can access the full set of independent parameters ($\Delta_1$, $\Delta_2$, and \DeltaT) using spectra with a pronounced low-energy \YSR state, which gives rise to well-resolved thermal resonances (these are caused by thermally excited quasiparticles tunneling between tip and sample)~\cite{Ruby2015b}. To this end, we used Mn adatoms, which show such low-energy \YSR resonances~\cite{Ruby2015b}.
The \YSR resonance and its thermal counterpart occur symmetric to \DeltaT at $\pm\left(\DeltaT+\varepsilon\right)$ and $\pm\left(\DeltaT-\varepsilon\right)$, respectively.  This allows us to determine \DeltaT unambiguously. Then, spectra of the pristine surface acquired with the same tip show clear BCS resonances at $\DeltaT+\Delta_{1,2}$ and we can unambiguously determine $\Delta_{1}$ and $\Delta_{2}$. Because these are bulk properties of the substrate, their energy can then serve to determine \DeltaT for every new tip. This procedure enables a reliable determination of the energies of subgap resonances in each Co chain.

\section{Cobalt chains without cluster termination}

Self-assembled cobalt chains grow on Pb(110) at elevated temperatures along the \dir{1\bar{1}0} direction. The majority of chains ($\approx 75\,\%$) emerge from a Co cluster and possess a single free end. We focused on these chains in the main text. The length of these chains is measured between the free end and the cluster onset. Approximately $25\,\%$ of the chains are either not connected to any cluster (although one end might fall on a step edge), or they are interlinking two clusters.

\begin{figure}[t]
	\includegraphics[width=\columnwidth]{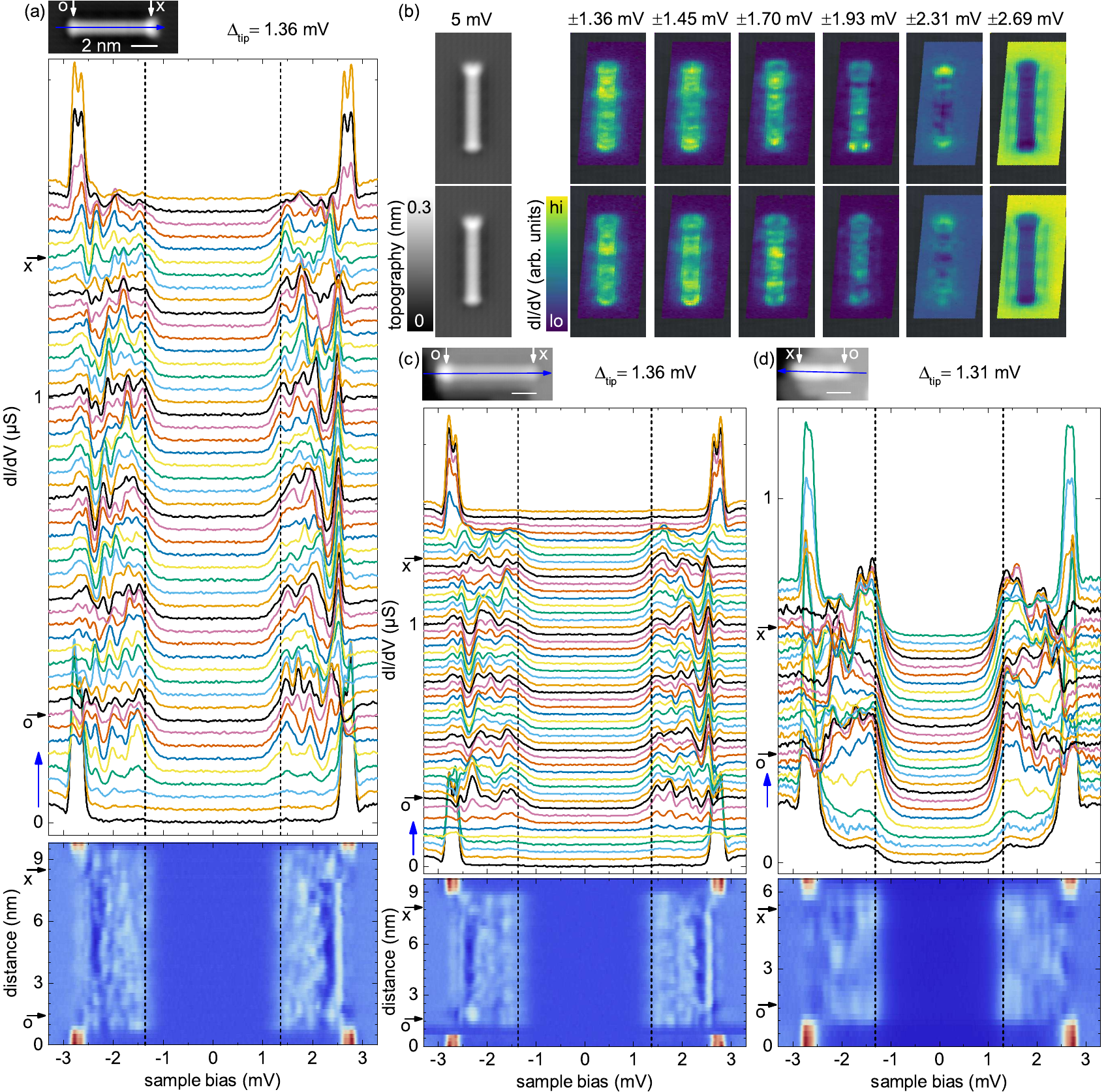}
	\caption{Three cobalt chains without a cluster at either end. The \didv spectra of the subgap structure in (a), (c), and (d) are acquired with a superconducting tip along the blue arrows sketched in the topography images on top of the spectra. The dashed lines mark the value of the tip gap. Zero energy excitations should appear here. A false color plot of the spectral intensity of all spectra is shown below each stacked graph. (b) shows \didv maps of the chain in (a) at selected energies. Feedback regulated in each point with setpoints $5\umV$, $200\upA$ (a--c); $4\umV$, $500\upA$ (d). Bias modulation: $15\umuV_\mathrm{rms}$. The spectra are offset for clarity by $30\unS$ (a), $50\unS$ (c), and $23\unS$ (d), respectively. The lateral distance between the spectra is $200\upm$ (a), $196\upm$ (c), and $225\upm$ (d), respectively. The chain lengths are $\simeq8.3\unm$ in (a) and (b), $7.5\unm$ in (c), and $4.1\unm$ in (d).}
	\label{Sfig:Chainwocluster}
\end{figure}

In \Fig{Chainwocluster}, we present three chains of different length not being connected to any cluster. The \didv spectra qualitatively exhibit the same features $\alpha$, $\beta$ (albeit different in details) as the spectra presented in the main text. 
Only the shortest chain [\Fig{Chainwocluster}(d)] shows a less pronounced \YSR resonance $\alpha$ at $\simeq2.5\umeV$. In all cases, the coherence peaks of the $s$-wave superconductor Pb are recovered within a few \AA ngstroms away from the chain ends. For none of the chains, we observe a localization of a zero-energy resonance at the chain ends. We rather point out that other resonance show indications of localization at the chain ends [e.g., see \didv maps at $\pm1.45$\,mV and $\pm2.31$\,mV in \Fig{Chainwocluster}(b)]. 

\section{Cobalt chains between two clusters}

\begin{figure}[t]
	\includegraphics[width=\columnwidth]{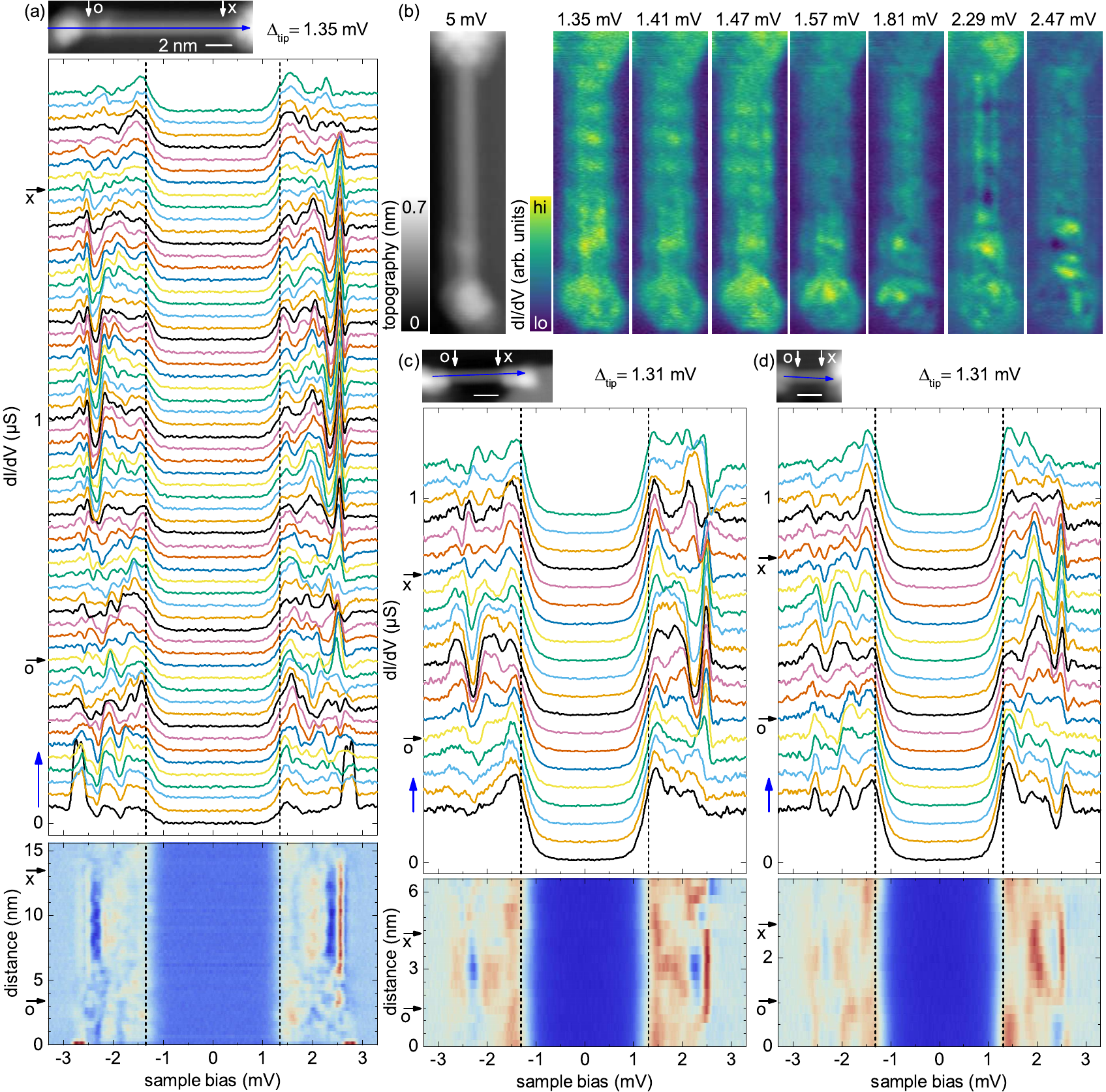}
	\caption{Three cobalt chains interlinking two clusters. The \didv spectra of the subgap structure in (a), (c), and (d) are acquired with a superconducting tip along the blue arrows sketched in the topography images on top of the spectra. The dashed lines mark the value of the tip gap. Zero energy excitations should appear here. A false color plot of the spectral intensity of all spectra is shown below each stacked graph. In (b), \didv maps at specific energies of the chain shown in (a) are plotted. The chain lengths are $\simeq11.5\unm$ in (a) and (b), $4.1\unm$ in (c), and $2.8\unm$ in (d). The setpoint is $5\umV$, $200\upA$ (a,b) and $4\umV$, $500\upA$ (c,d). The spectra are offset for clarity by $30\unS$ (a), and $50\unS$ (c,d). The lateral distance between the spectra is $265\upm$ (a), $343\upm$ (c), and $200\upm$  (d), respectively. The bias modulation is $15\umuV_\mathrm{rms}$ for spectra (a,c,d), and $25\umuV_\mathrm{rms}$ for the \didv-maps (b).}
	\label{Sfig:Chaininterlinkcluster}
\end{figure}

\Figure{Chaininterlinkcluster} shows \didv spectra of three chains, which emerge between two clusters. All chains exhibit a rich subgap structure similar to the chains conntected to only one cluster or no cluster termination. Again, we observe no signatures of the localization of a state at zero energy.

\newpage

\section{Additional data on the $\textbf{Co}$ chains presented in the main text}

In Fig.~3 of the main text, we presented \didv data on a chain of $10.3\unm$ length.
For completeness, here we show additional data on the same chain. 
\Figure{SFigChain4Fig3OfMainText} presents the full set of \didv spectra recorded along the central axis of the chain.

\begin{figure}[hbt]
	\includegraphics[width=\columnwidth]{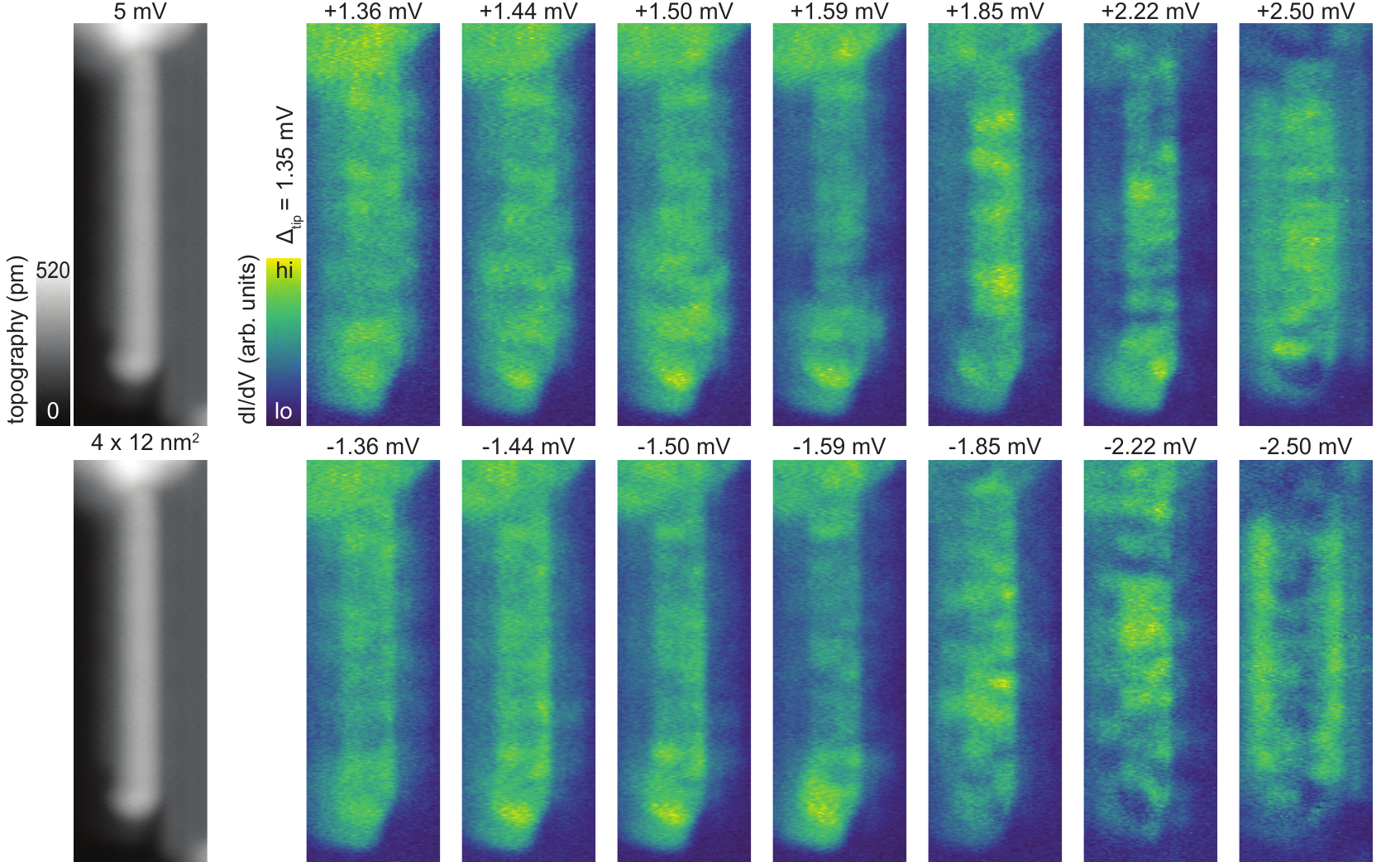}
	\caption{\didv maps of the Co chain on Pb(110) shown in Fig.~3 of the main text. The chain length is $\simeq10.3\unm$. The maps were recorded with a superconducting tip with $\DeltaT=1.35\umeV$. Note, that the energy resolution is $\simeq60\umueV$. Setpoint: $5\umV$, $200\upA$. Bias modulation: $25\umuV_\mathrm{rms}$.}
	\label{Sfig:SuppFigdIdVMaps}
\end{figure}

\begin{figure}[h]
	\includegraphics[width=\columnwidth]{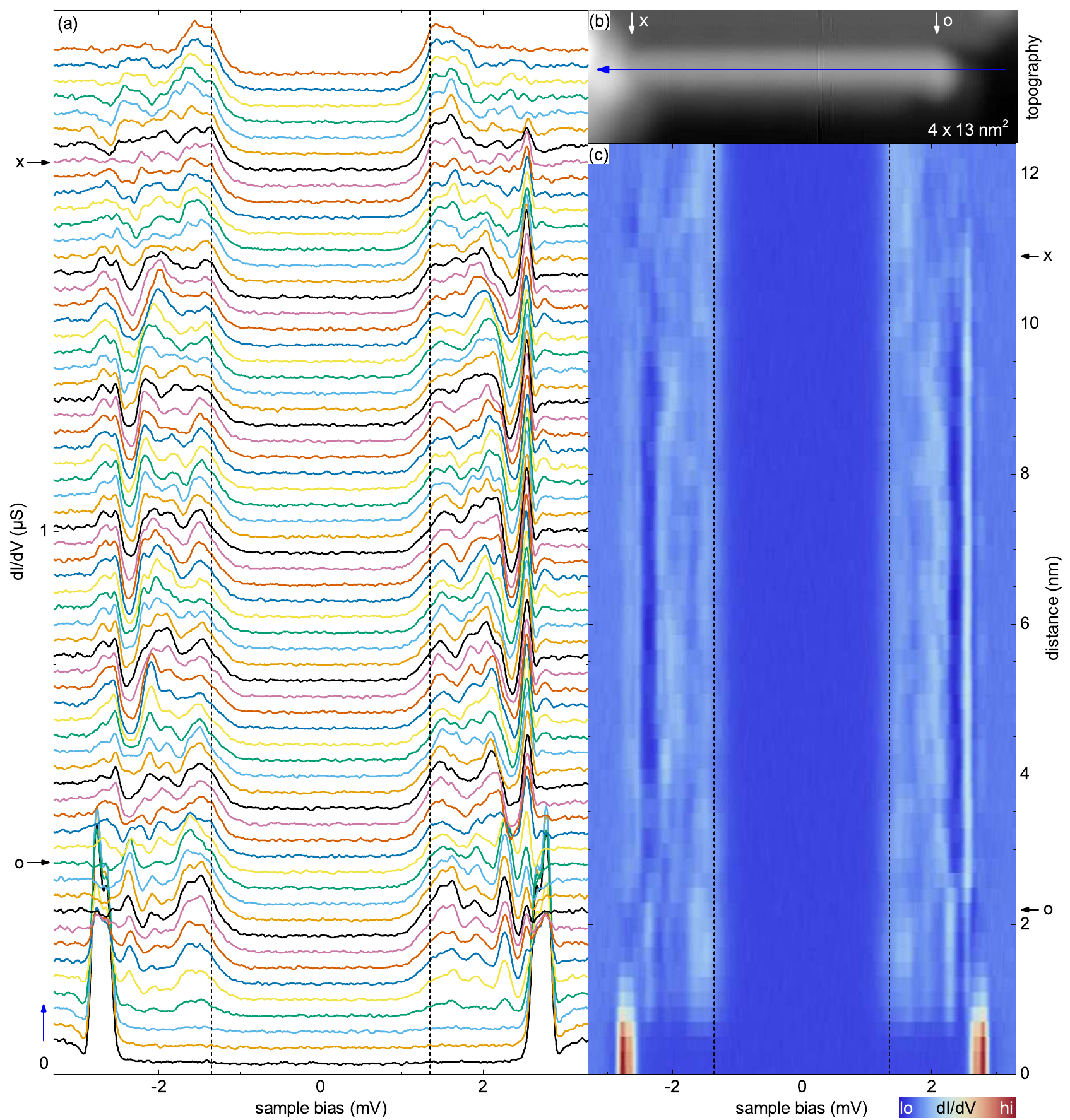}
	\caption{(a) Full set of \didv spectra recorded at the Co chain on Pb(110) shown in (b) along the blue arrow. The chain is the same as the one in Fig.~3 of the main text. The chain length is $\simeq10.3\unm$. The spectra are offset for clarity by $30\unS$. The lateral distance between the spectra is $200\upm$. Tip gap: $1.35\umeV$. Setpoint: $5\umV$, $200\upA$. A false color plot of the spectra is shown in (c).}
	\label{Sfig:SFigChain4Fig3OfMainText}
\end{figure}

To image the possible localization of Majorana end states, we recorded \didv maps (\Fig{SuppFigdIdVMaps}) with a superconducting tip with $\pm\DeltaT\simeq\pm1.35\umV$. The \didv maps at different subgap energies show variations in the intensity along the chain. We observe an increased intensity at the chain end for low-lying \YSR states at $\pm1.59$ and $\pm1.50$\,mV. The tail of these resonances spreads down to zero energy and is likely at the origin of the faint contrast increase at the chain end in the maps at $\pm1.36$\,mV (equal to $\pm\DeltaT$ within the energy resolution).

\clearpage

\begin{figure}[bth]
	\includegraphics[width=\columnwidth]{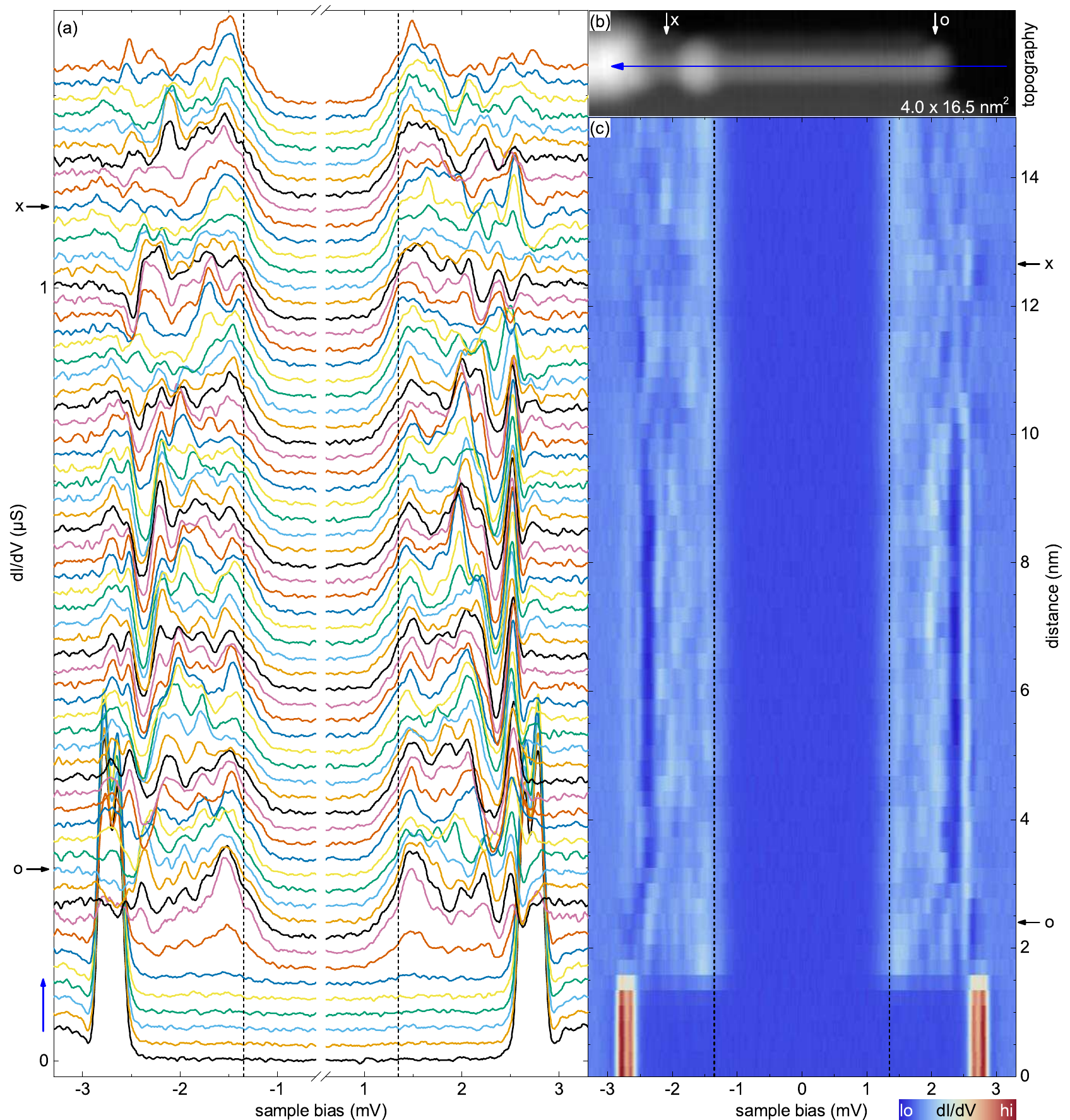}
	\caption{(a) Set of \didv spectra recorded at the Co chain on Pb(110) shown in (b) along the blue arrow. The chain is the same as the one in Fig.~4(a) of the main text. The chain length is $\simeq11.7\unm$. The spectra are offset for clarity by $20\unS$. The lateral distance between the spectra is $241\upm$. Tip gap: $1.35\umeV$. Setpoint: $5\umV$, $200\upA$. A false color plot of the spectral intensity of all spectra with respect to the lateral distance is shown in (c).}
	\label{Sfig:SFigChain5Fig4aOfMainText}
\end{figure}

\begin{figure}[bth]
	\includegraphics[width=\columnwidth]{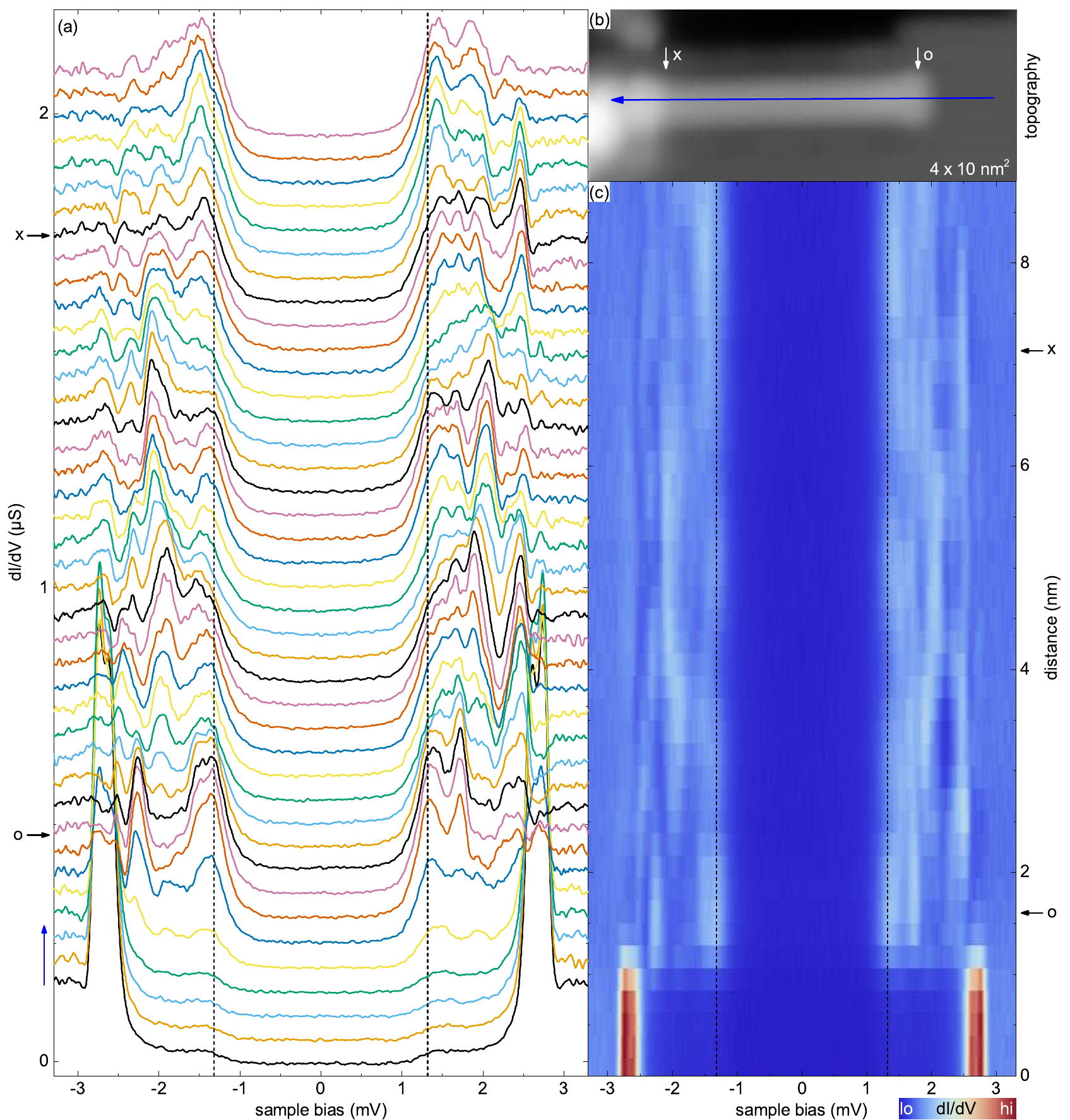}
	\caption{(a) Set of \didv spectra recorded at the Co chain on Pb(110) shown in (b) along the blue arrow. The chain is the same as the one in Fig.~4(b) of the main text. The chain length is $\simeq6.2\unm$. The spectra are offset for clarity by $50\unS$. The lateral distance between the spectra is $225\upm$. Tip gap: $1.32\umeV$. Setpoint: $4\umV$, $500\upA$. A false color plot of the spectral intensity of all spectra with respect to the lateral distance is shown in (c).}
	\label{Sfig:SFigChain5Fig4bOfMainText}
\end{figure}

\begin{figure}[bth]
	\includegraphics[width=\columnwidth]{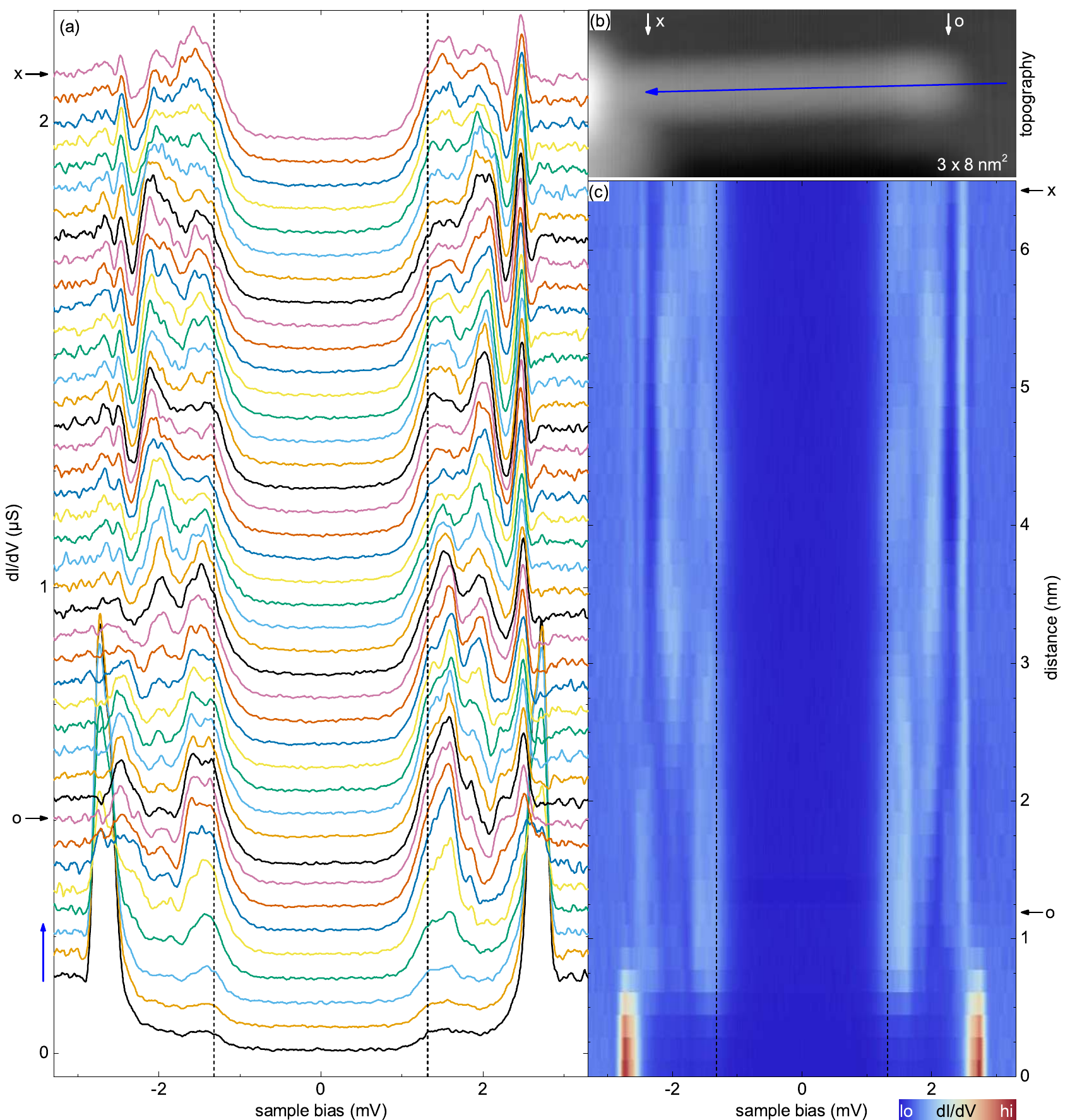}
	\caption{(a) Set of \didv spectra recorded at the Co chain on Pb(110) shown in (b) along the blue arrow. The chain is the same as the one in Fig.~4(c) of the main text. The chain length is $\simeq6.1\unm$. The spectra are offset for clarity by $50\unS$. The lateral distance between the spectra is $167\upm$. Tip gap: $1.32\umeV$. Setpoint: $4\umV$, $500\upA$. A false color plot of the spectral intensity of all spectra with respect to the lateral distance is shown in (c).}
	\label{Sfig:SFigChain5Fig4cOfMainText}
\end{figure}

In Fig.~4 of the main text, we show selected \didv spectra of four different Co chains on Pb(110). In \Fig{SFigChain5Fig4aOfMainText}, \Fig{SFigChain5Fig4bOfMainText} and \Fig{SFigChain5Fig4cOfMainText}, we present the full sets of \didv spectra recorded along the central axis of the chains in Fig.~4(a), (b), and (c), respectively. All characteristics described in the main text are observed in these chains.

\clearpage

\section{Tight binding band structure of a cobalt chain}

\begin{table}[bth]
  \caption{Slater-Koster tight-binding model parameters for Co (in $\ueV$). The spin-polarized orbital energies are denoted as $\epsilon$. The hopping integral values $V$ are calculated for the nearest-neighbor distance of bulk Co (hcp) with $a=2.486\uAA$.}
    \begin{tabular}{ccc}
      \hline 
      Parameters & Spin up & Spin down\tabularnewline
      \hline 
      $\epsilon(d_{z^{2}})$ & 7.175  & 8.996 \tabularnewline
      $\epsilon(d_{yz,xz})$ & 7.156  & 8.973 \tabularnewline
      $\epsilon(d_{xy,x^{2}-y^{2}})$ & 7.197 & 9.011 \tabularnewline
      $V(dd\sigma)$ & -0.528  & -0.599 \tabularnewline
      $V(dd\pi)$ & 0.325  & 0.376 \tabularnewline
      $V(dd\delta)$ & -0.052  & -0.060 \tabularnewline
      \hline 
    \end{tabular}
		\label{tab:TBModel}
\end{table}
In Fig.~5 of the main text, we show the band structure of a linear suspended cobalt chain, which is calculated using a tight binding model. The chain is aligned along the  $\hat{\mathbf{z}}$ direction. The parameters are taken from Ref.~\cite{handbook}, and are given in Tab.~\ref{tab:TBModel}. The calculations first consider the $d$ orbitals without spin-orbit coupling or superconductivity [Fig.~5(a) of the main text]. For each spin orientation, we obtain two doubly-degenerate bands from the \dorb{xz,yz} and \dorb{xy,x^2-y^2} orbitals, and a non-degenerate band from the \dorb{z^2} orbital. Notice that the band structure is very similar to the one of iron chains on Pb(110) \cite{Nadj14}.

A weak spin-orbit coupling lifts the degeneracy of the bands. We take this into account by adding the spin-orbit coupling Hamiltonian
\begin{equation}
  H_{\rm so} = \lambda_{\rm so} \mathbf{L} \cdot \mathbf{s},
\end{equation}
where $\lambda_{\rm so}$ denotes the strength of the spin-orbit coupling, $\mathbf{L}$ is the orbital angular momentum of the electron and $\mathbf{s}$ is the spin angular momentum. The angle between the magnetization direction $\braket{\mathbf{s}}$ and $\hat{\mathbf{z}}$ is denoted as $\theta$. The band structure with spin-orbit coupling for $\theta=\pi/2$ is shown in Fig.~5(b) of the main text. We use $\lambda_\mathrm{so}=0.2\ueV$, which is the same as the value for Fe in Ref.~\cite{Li14}. If the magnetization is not perpendicular to the chain direction, the bands are further split. For $\theta=2\pi/5$ and $\lambda_\mathrm{so}=0.2\ueV$, the band structure is shown in Fig.~5(c) of the main text.

\section{Discussion of the band filling of cobalt chains on Pb(110)}

The Fermi energy is a crucial property of the system, as it determines the number Fermi points within the band structure. Here we discuss two scenarios for a 1D Co chain, which result in different positions of the Fermi level.

In the main text, we discussed a scenario which is deduced from the single-atom limit. When a single Co atom is deposited onto a Pb surface, electrons can be transferred into the bulk Pb, and the Co atom becomes positively charged, presumably between $\mathrm{Co}^{2+}$ and $\mathrm{Co}^{3+}$, the most common oxidation states of cobalt. This empties the $s$ levels and partially the $d$ levels. Thus, the number of valence ($d$) electrons per Co is between $6$ and $7$. We consider this a likely scenario and, hence, use it for the determination of the chemical potential in Figs.~5(b) and (c) of the main text.

\begin{figure}[bth]
	\includegraphics[width=0.75\columnwidth]{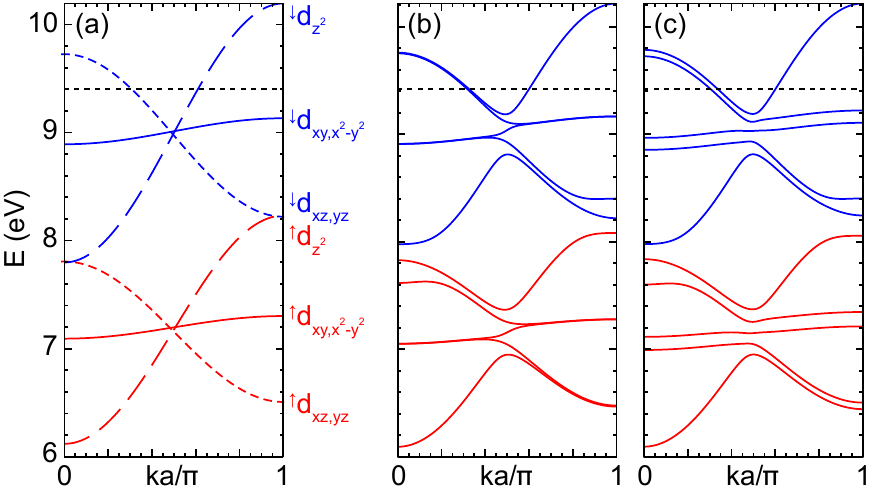}
	\caption{Tight-binding band structure of a linear suspended Co chain with interatomic distance $a=2.486\uAA$ neglecting (a), or including (b,c) spin-orbit coupling, respectively. Spin-orbit coupling parameter: $\lambda_\mathrm{so}=0.2\ueV$. The angle between the magnetization and the chain direction is $\pi/2$ (b) and $2\pi/5$ (c), respectively. The splitting between the minority (\textit{blue}) and majority (\textit{red}) bands is due to the exchange interaction. The horizontal black lines indicate a possible value of the chemical potential when considering a scenario with nine $d$ electrons. This would predict three Fermi points and robust topological superconductivity.}
	\label{Sfig:BandStructure2ndModel}
\end{figure}

A second scenario (see also \Refs{Li14,Nadj14}) is sketched for a Co chain in \Fig{BandStructure2ndModel}. Here, nine valence electrons per Co atom are assumed, which
corresponds to the number of valence electrons of a neutral Co atom. All nine electrons per atom presumably fill the $d$ bands, because the $s$-derived bands lie higher in energy for the 1D chains. In this scenario, the Co chains always have three Fermi points in the band structure [\Fig{BandStructure2ndModel}(c)], and would be expected to exhibit robust topological superconductivity.

\end{document}